\documentclass[reprint,amsmath,amssymb,aps,notitlepage,pre]{revtex4-2}

\usepackage{enumitem}
\usepackage{placeins} 
\usepackage{url}
\usepackage[hyperindex,breaklinks]{hyperref}
\hypersetup{linkcolor=blue, citecolor=red, colorlinks=true}
\usepackage{capt-of}
\usepackage{graphicx}
\usepackage{dcolumn}
\usepackage{bm}
 \usepackage{booktabs} 
\usepackage{array}
\usepackage{physics}

\begin{document}

\title{Tricriticality and chaos in a generalized Allee-logistic map}

\author{Marcelo A. Pires$^{1}$}
\thanks{piresma@cbpf.br}

\author{Jos{\'e}  S. Andrade Jr.$^{2}$}

\author{Hans J. Herrmann$^{2,3}$}

\affiliation{$^{1}$Centro Brasileiro de Pesquisas F\'{\i}sicas, Rio de Janeiro - RJ, 22290-180, Brasil \\ 
$^{2}$Universidade Federal do Cear{\'a}, Campus do Pici, 60451-970, Fortaleza - CE, Brasil \\
$^{3}$PMMH, ESPCI, 7 quai St. Bernard, 75005 Paris, France
}

\begin{abstract} 
We present a novel nonlinear dynamical model, the generalized Allee-logistic (GAL) map given by $x_{t+1} = r x_t (1 - x_t) G(x_t)$ where $G(x_t) = m (x_t - h) + 1 - m$   incorporates the Allee effect with magnitude $m$ and threshold $h$. 
The case $m = 0$ yields the logistic map with a continuous transition to extinction. Conversely, $m = 1$ 
recovers a previously studied model that undergoes only a discontinuous extinction-to-active transition. Between these extremes, the GAL map exhibits nontrivial phenomena, 
including tricriticality with a closed-form expression for the tricritical point and a universal crossover function.
Under a small external input, we verify Widom-like relations.
We also note that the Allee effect disfavors the onset of chaos.
Our work establishes additional bridges between analytically tractable chaotic maps, 
nonequilibrium tricriticality, and Allee effects. 
\end{abstract}
 
\maketitle

\section{Introduction}\label{sec:intro}

Nonlinear dynamical maps serve as foundational models for understanding criticality and universality in nonequilibrium systems. Investigations into such maps have revealed deep connections 
between bifurcations and nonequilibrium phase transitions~\cite{feigenbaum1978quantitative,chang1981iterative,kapustina2001scaling,kuznetsov2001two,ambika2002critical,xu2003experimental,kuznetsov2006effect,stynes2010scaling,teixeira2015convergence,girardi2019comment,corral2018finite,corral2025universal,martin2025finite,polli2025power}. Among these systems, the logistic map~\cite{may1976simple,ausloos2006logistic} stands as a  
canonical model in nonlinear dynamics
that exhibits a variety of phenomena including a continuous transition to extinction, a bifurcation cascade, intermittency and chaos.

Despite a rich phenomenology, the standard logistic map does not incorporate other ecological features such as  
Allee effects, which play a significant role in many areas of biology~\cite{korolev2014turning,sewalt2016influences,sun2016mathematical,amarasekare1998allee,fowler2002population,nath2022refugia,berec2008models,dos2015models,pires2019optimal,pires2022randomness,aleixo2009populational,leonel2019allee,el2023allee}. This effect manifests as a reduction in effective population growth, a phenomenon that can occur due to various reasons, such as difficulty in finding mates, reduced cooperative behaviors, and other factors~\cite{courchamp2008allee,tobin2011exploiting}.
Empirical evidence for the Allee effect is observed in  terrestrial arthropods, aquatic invertebrates, mammals, birds, fish, and reptiles~\cite{kramer2009evidence,fauvergue2013review,angulo2018allee,branco2024widespread} as well as engineered bacterial populations~\cite{smith2014programmed}.

Previous research~\cite{jung2020chaotic} introduced the Allee effect into the logistic map and identified discontinuous extinction-to-active transitions. 
We move beyond this model by proposing the Generalized Allee-Logistic (GAL) map that exhibits both continuous and discontinuous transitions to extinction. 

Tricriticality in nonequilibrium systems has been investigated across diverse disciplines such as epidemiology~\cite{janssen2004generalized,pires2023tricritical}, 
ecology~\cite{bhattacharyya2024emergence,windus2008cluster},
turbulence~\cite{jayasingh2025tricritical},
percolation~\cite{araujo2011tricritical,cellai2011tricritical} and 
directed percolation \cite{lubeck2006tricritical,grassberger2006tricritical}. The GAL map, with its analytical tractability, offers a versatile tool to explore critical and tricritical features in chaotic dynamics, linking ecological Allee effects to broader nonequilibrium phase transition theory.

\begin{figure*}[!htb]
 \centering
 \includegraphics[width=0.99\textwidth]{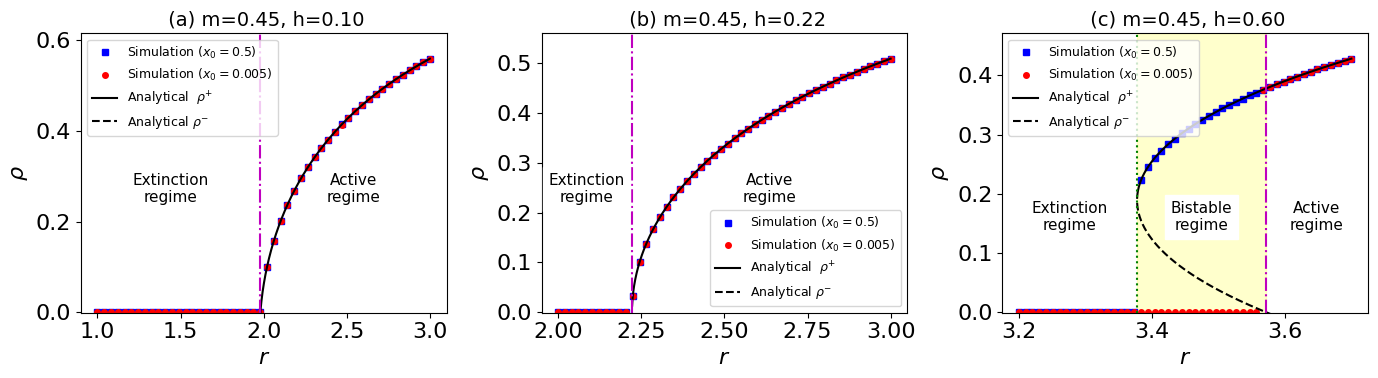}  
\caption{
 Order parameter $\rho$, Eq.~\eqref{eq:rho}, versus the control parameter $r$. Transitions in the GAL map: (a) continuous, $h<h_T$; (b) tricritical, $h=h_T$; (c) discontinuous with bistability between $r_b$ and $r_c$, $h>h_T$. 
The analytical results given by Eqs.~(\ref{eq:sol3_0}, \ref{eq:sol3_Iplus}, \ref{eq:rc}, \ref{eq:tcp}, \ref{eq:rb}) agree well with simulations from different initial conditions. 
} 
\label{fig:log_allee_transitions} 

\end{figure*}

The paper is organized as follows: Section~\ref{sec:model} presents our new model, Section~\ref{sec:results} analyzes its properties, and Section~\ref{sec:finalremarks} provides our concluding remarks.

\section{Model}\label{sec:model}

We introduce a generalized Allee-logistic  (GAL) map given by
\begin{align} \label{eq:gal_map1}
x_{t+1} &= r x_t (1 - x_t) G(x_t), \\
G(x_t) &= m (x_t - h) + 1 - m,
\end{align}
where $x_t$ represents the population density at time $t = \{0,1,2,\ldots\}$, $m \in [0, 1]$ is the magnitude of the Allee effect, $h \in [0, 1]$ is the Allee threshold. The parameter $r$ is the intrinsic growth rate of the population.
Since  $x_t - h \leq 1$, the factor $G(x_t) \leq 1$  incorporates the Allee effect through a reduction in the effective population growth. 
Clearly, in Eq.(\ref{eq:gal_map1}), only the parameters that allow the population fraction to satisfy 
$x_t \in [0, 1]$ are permitted.

Observe that when:
\begin{itemize}
 \item $m = 0$, we find $G(x_t) = 1$, and the model reduces to the standard logistic map.
 \item $m = 1$, we obtain $G(x_t) = x_t - h$, and the model recovers a previous  particular case~\cite{jung2020chaotic}.
\end{itemize}

\section{Results}\label{sec:results}

The GAL map can be written as 
\begin{align}
 x_{t+1} = f(x_t) = A_{r,m} x_t^3 + B_{r,m,h} x_t^2 + C_{r,m,h} x_t, \label{eq:gal_map}
\end{align}
where the coefficients are given by:
\begin{align}
A_{r,m} &= -r m, \label{eq:A_coeff} \\
B_{r,m,h} &= r(2m + m h - 1), \label{eq:B_coeff} \\
C_{r,m,h} &= r(1 - m - m h). \label{eq:C_coeff}
\end{align}

From now on, to avoid overloading the notation, we will omit the subscripts in Eqs.~(\ref{eq:A_coeff}-\ref{eq:C_coeff}).

\subsection{Stationary state}
The condition for the stationary state is
\begin{align}
x_{t} = x_{t+1} = x_{\infty} \equiv \rho. \label{eq:rho}
\end{align}
Then, using Eq.~\eqref{eq:gal_map}, we obtain
\begin{align}
    f(\rho) - \rho = A \rho^3 + B \rho^2 + (C - 1) \rho = 0. \label{eq:steady_state}
\end{align}

The three solutions of Eq.~\eqref{eq:steady_state} are:
\begin{align} 
\rho_0 &= 0, \label{eq:sol3_0} \\
\rho^{\pm} &= 
\frac{-B \pm \sqrt{B^2 - 4A(C-1)}}{2A}. \label{eq:sol3_Iplus}
\end{align}

The trivial fixed point $\rho_0 = 0$ represents an extinction state, and the nontrivial solutions $\rho^{\pm}$ represent an active population. These analytical results and corresponding simulations are displayed in Fig.~\ref{fig:log_allee_transitions}.

\begin{figure*}[!htb]
 \centering
 \includegraphics[width=0.99\textwidth]{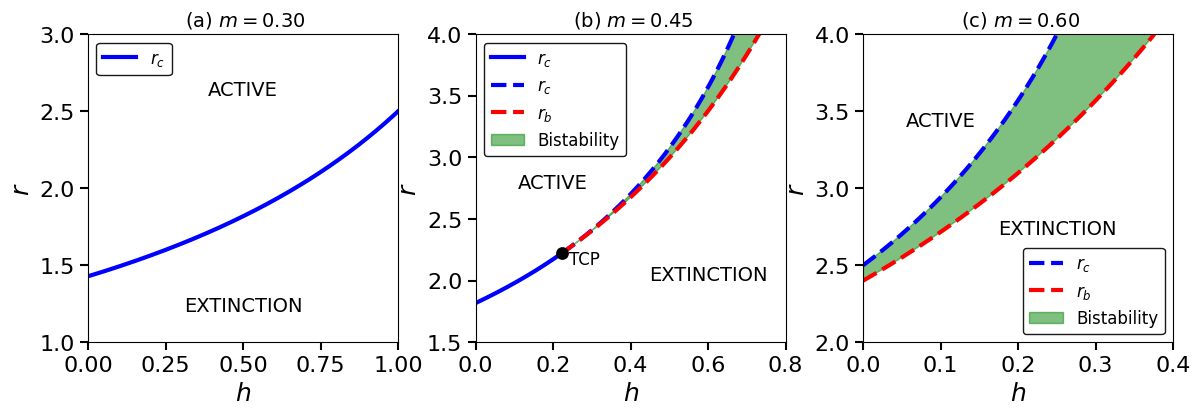}
\caption{  Phase diagram of the GAL map in the conditions of 
Eq.~\ref{eq:TCP_cond1} (panel a),
Eq.~\ref{eq:TCP_cond2} (panel b), and 
Eq.~\ref{eq:TCP_cond3} (panel c). 
Solid and dashed curves represent continuous and discontinuous transitions, respectively. The green shaded region represents the bistable regime. The TCP emerges at the intersection of continuous and discontinuous transition boundaries.} 
\label{fig:log_allee_tcp}
\end{figure*}

\subsection{Critical point}
To analyze stability, we first compute the derivative of the map:
\begin{align}
    f'(x) = 3A x^2 + 2B x + C. \label{eq:derivative}
\end{align}

The critical condition, $|f'(0)| = 1$, combined with  Eq.~\eqref{eq:C_coeff} yields 
\begin{align}
    r_c = \frac{1}{1 - m - m h}. \label{eq:rc}
\end{align}

This equation gives the critical point where the extinction state loses stability, as evident in Fig.~\ref{fig:log_allee_transitions}.

\subsection{Tricritical point (TCP)}

The TCP marks the point at which continuous and discontinuous transition solutions meet.
This requires the coalescence of $\rho^{\pm}$ with $\rho_0$.
Consequently, the coefficients of the linear and quadratic terms in Eq.~(\ref{eq:steady_state}) must vanish at the transition:
\begin{align}
    C_T = 1, \quad B_T = 0. \label{eq:tcp_B_C_condition}
\end{align}
Substituting these conditions in Eqs.~(\ref{eq:B_coeff}, \ref{eq:C_coeff}) leads to a closed-form for the TCP:
\begin{align}
    (h_T, r_T) = \left( \frac{1 - 2m}{m}, \frac{1}{m} \right). \label{eq:tcp}
\end{align}
The condition $0\leq h_T\leq1$ with $h_T=(1 - 2m)/m$ implies that a TCP exists if $1/3 \leq m\leq 1/2$. The transition to extinction can be:
\begin{align}
& \text{Only continuous if } m<1/3 \label{eq:TCP_cond1} \\ 
& \text{Continuous or discontinuous if } 1/3 \leq m\leq 1/2 \label{eq:TCP_cond2} \\ 
& \text{Only discontinuous if } m>1/2 \label{eq:TCP_cond3} 
\end{align}
The above scenarios are illustrated in Fig.~\ref{fig:log_allee_tcp}.

\begin{figure}[!htb] \centering
 \includegraphics[width=0.432\textwidth]{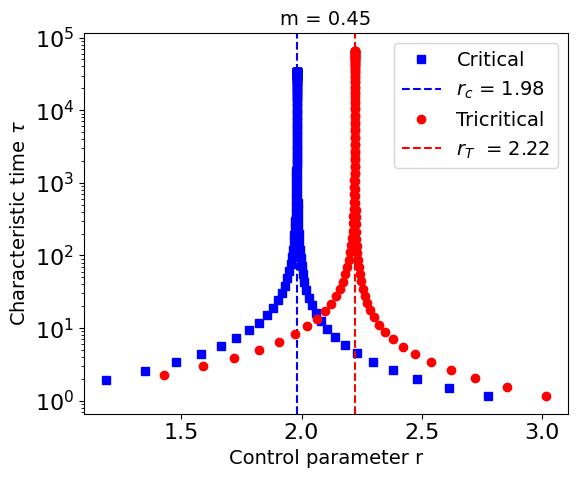}
\caption{Relaxation dynamics with characteristic time $\tau$ near critical ($r_c$, blue) and tricritical ($r_T$, red) points. } 
\label{fig:log_allee_divergence}
\end{figure}

\begin{figure*}[!htb]
\centering

\includegraphics[width=0.321\textwidth]{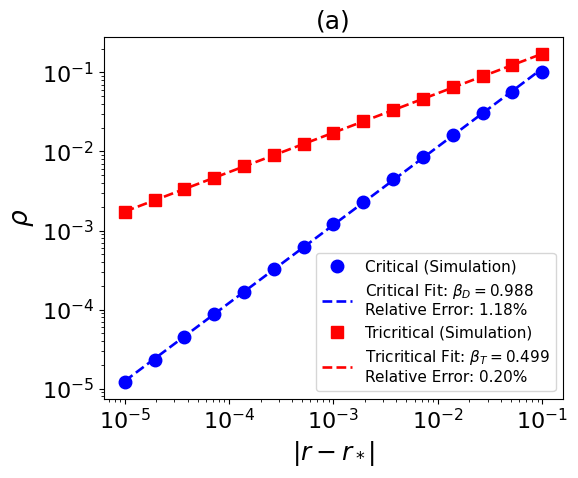}
\includegraphics[width=0.321\textwidth]{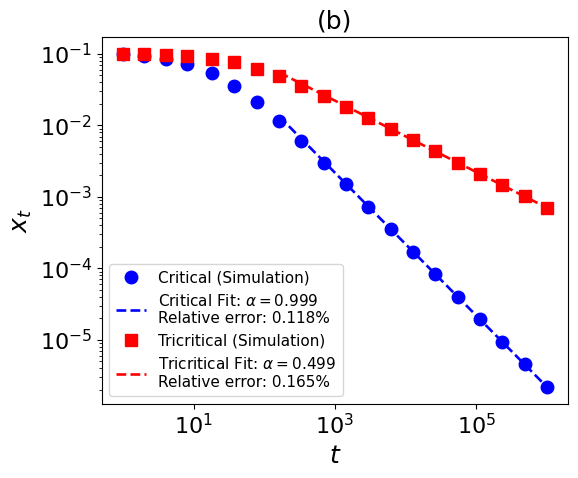}
\includegraphics[width=0.321\textwidth]{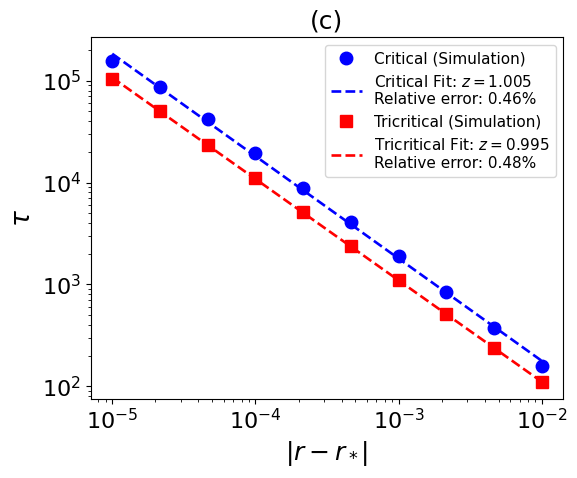}

\includegraphics[width=0.321\textwidth]{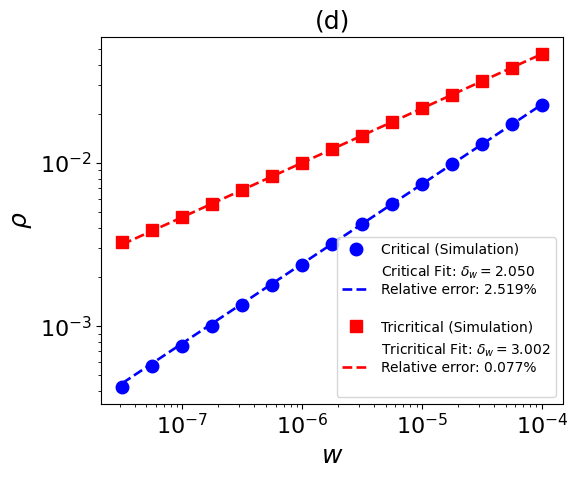}
\includegraphics[width=0.321\textwidth]{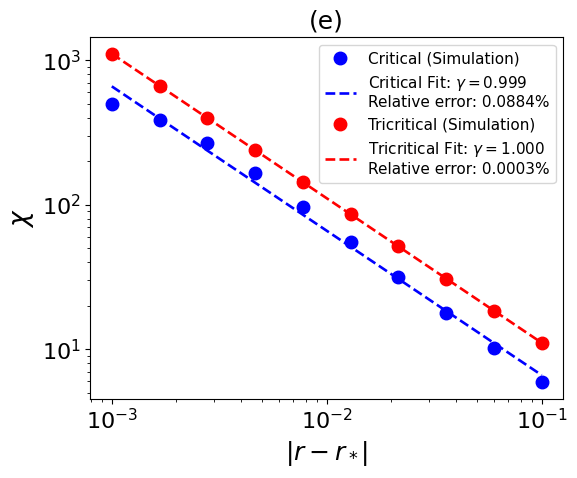}
\includegraphics[width=0.321\textwidth]{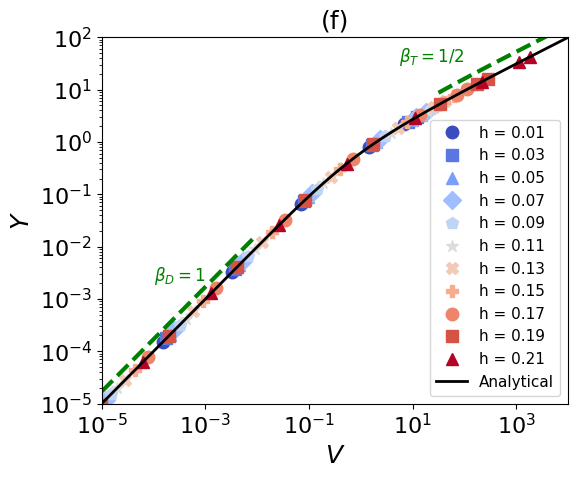}

\caption{ 
Scaling of the GAL map. All panels are in log-log scale. For the critical case $r_{*}=r_c$ and for the  tricritical case $r_{*}=r_T$. 
In panels (a-e) we show the fits with the estimated exponents. 
(f) Data collapse for the GAL map showing the scaled order parameter $Y$ versus scaled control parameter $V$. The crossover is    in accordance with Eq.~\eqref{eq:universal_function}. 
The numerical results are in good agreement with the theoretical results summarized in Table~\ref{tab:comparison_DP_TDP}.
}
\label{fig:log_allee_scaling}
\end{figure*}

\subsection{Bistability threshold}
When the two nontrivial fixed points coincide ($\rho_- = \rho_+$), the discriminant of Eq.~\eqref{eq:sol3_Iplus} must vanish. Solving $\Delta = 0$ yields:
\begin{align}
    r_b = \frac{4m}{(1 - m h)^2}. \label{eq:rb}
\end{align}

From Eqs.~(\ref{eq:rc}, \ref{eq:rb}) we have the thresholds for the region where the system has two distinct stable fixed points, and the long-term dynamics depends on the initial condition, as shown in Figs.~(\ref{fig:log_allee_transitions}-\ref{fig:log_allee_tcp}).

\subsection{Temporal scaling}

We turn our focus to the relaxation toward the fixed points. 
Following~\cite{de2013relaxation}, we approximate $ x_{t+1} - x_t = \frac{x_{t+1} - x_t}{(t+1)-(t)} = \frac{\Delta x}{\Delta t} \approx \frac{dx}{dt}$ in the limit of small perturbations near the extinction state $\rho_0 = 0$. Then, we rewrite Eq.~\eqref{eq:gal_map} as:
\begin{align}
\frac{dx}{dt} \approx A x_t^3 + B x_t^2 + (C-1) x_t. 
  \label{eq:gal_map2}
\end{align}

\subsubsection{Above the extinction point}

When $C>1$, the linear term dominates in both the critical and tricritical cases. Thus, 
from Eq.~\eqref{eq:gal_map2}
we obtain an exponential behavior $x_t \sim e^{t/\tau}$   which leads to:
\begin{align}
\tau &\sim |r - r_c|^{-z_D}, \quad z_D=1, \label{eq:nu_D_above} \\
\tau &\sim |r - r_T|^{-z_T}, \quad z_T=1. 
\label{eq:nu_T_above} 
\end{align}
These Eqs.~(\ref{eq:nu_D_above}, \ref{eq:nu_T_above}) indicate  that the characteristic time has a divergence-like behavior, as confirmed in Fig.~\ref{fig:log_allee_divergence}.


\subsubsection{At the extinction point}

\paragraph{Critical case} 
When $B \neq 0$ and $C \to 1$, the dominant term in  Eq.~\eqref{eq:gal_map2} is   $x^2$, which gives us $\dot{x} \approx B x^2$. Integrating this expression yields
\begin{align}
x(t) &\sim t^{-\alpha_D}, \quad \text{with} \quad \alpha_D = 1. \label{eq:delta_D}
\end{align}

\paragraph{Tricritical case} 
When $B \to 0$ and $C \to 1$, the dominant term in  Eq.~\eqref{eq:gal_map2} becomes  $x^3$, resulting in 
$\dot{x} \approx A x^3$. Integration leads to the scaling relation:
\begin{align}
x(t) &\sim t^{-\alpha_T}, \quad \text{with} \quad \alpha_T = \frac{1}{2}. \label{eq:delta_T}
\end{align}

Therefore, at the transition we note different exponents given by Eqs.~(\ref{eq:delta_D}, \ref{eq:delta_T}), as shown in Fig.~\ref{fig:log_allee_scaling}.

\begin{table*}[!htb]
\caption{\label{tab:comparison_DP_TDP}
Scaling for the GAL map, showing correspondence with mean-field (MF) directed percolation (DP) and tricritical directed percolation (TDP) universality classes~\cite{lubeck2006tricritical,grassberger2006tricritical}. 
The Widom-like relations are satisfactorily verified. 
The simulations in Figs.~(\ref{fig:log_allee_scaling}, \ref{fig:log_allee_lyapunov}) confirm the theoretical predictions summarized in this table.
}
\begin{ruledtabular}
\begin{tabular}{lll}
Property & Criticality in the GAL map (MF-DP) & Tricriticality in the GAL map (MF-TDP) \\
\hline
Characteristic relaxation time 
& $\tau \sim |r - r_c|^{-z_D}, \quad z_D = 1$, Eq.~\eqref{eq:nu_D_above} 
& $\tau \sim |r - r_T|^{-z_T}, \quad z_T = 1$, Eq.~\eqref{eq:nu_T_above} \\
Temporal decay at transition 
& $x_t(r_c) \sim t^{-\alpha_D}, \quad \alpha_D = 1$, Eq.~\eqref{eq:delta_D}
& $x_t(r_T) \sim t^{-\alpha_T}, \quad \alpha_T = 1/2$, Eq.~\eqref{eq:delta_T} \\
Order parameter & 
$\rho \sim |r - r_c|^{\beta_D}, \quad \beta_D = 1$, Eq.~\eqref{eq:betaD} 
& $\rho \sim |r - r_T|^{\beta_T}, \quad \beta_T = 1/2$, Eq.~\eqref{eq:betaT} \\
External flux scaling at transition 
& $\rho(r_c) \sim w^{1/\delta_D^w}, \quad \delta_D^w = 2$, Eq.~\eqref{eq:deltaD}
& $\rho(r_T) \sim w^{1/\delta_T^w}, \quad \delta_T^w = 3$, Eq.~\eqref{eq:delta_wT} \\
Dynamical susceptibility 
& $\chi \sim |r - r_c|^{-\gamma_D}, \quad \gamma_D = 1$, Eq.~\eqref{eq:gammaD}
& $\chi \sim |r - r_T|^{-\gamma_T}, \quad \gamma_T = 1$, Eq.~\eqref{eq:gammaT}  \\
FTLE temporal relaxation 
& $\lambda_t(r_c) \sim t^{-\theta_D} \ln t, \quad \theta_D = 1$, Eq.~\eqref{eq:lambda_critical_scaling}
& $\lambda_t(r_T) \sim t^{-\theta_T} \ln t, \quad \theta_T = 1$, Eq.~\eqref{eq:lambda_tricritical_scaling} \\
Widom-like relation & $\delta_D^w - \frac{\gamma_D}{\beta_D} = 1$ & $\delta_T^w - \frac{\gamma_T}{\beta_T} = 1$ \\
\end{tabular}
\end{ruledtabular}
\end{table*}

\begin{figure}[!htb] 
\centering
 \includegraphics[width=0.444\textwidth]{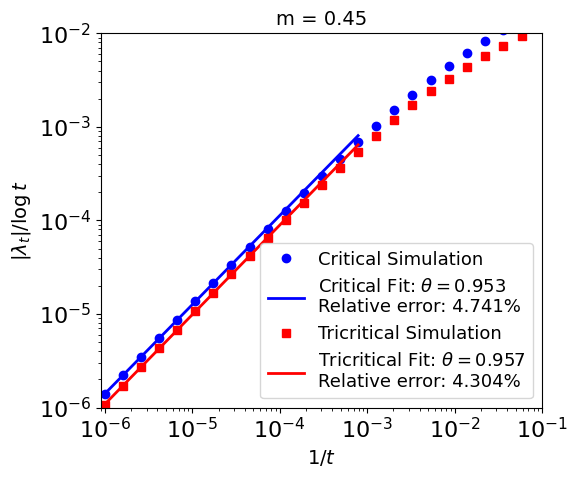}
\caption{
Scaling of the decay of the scaled modular FTLE versus $t^{-1}$ with variables given by Eqs.~(\ref{eq:lambda_critical_scaling}, \ref{eq:lambda_tricritical_scaling}). Representative parameters: $h=0.1$ (critical) and $h=h_T$ (tricritical).
} 
\label{fig:log_allee_lyapunov}
\end{figure}

\begin{figure}[!htb]
\centering
\includegraphics[width=0.444\textwidth]{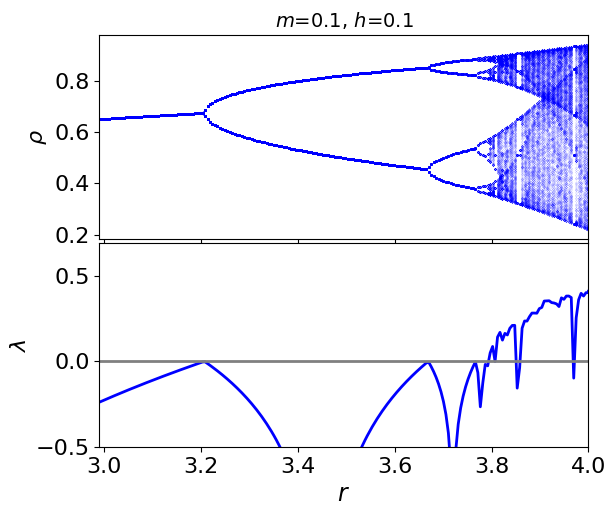}
\caption{Bifurcation diagram (top) and Lyapunov exponent $\lambda$ (bottom) for the GAL map for representative parameters.}
\label{fig:log_allee_bifur_lyap}
\end{figure}

\subsection{Scaling of the order parameter}

From Eq.~\eqref{eq:steady_state}, we derive  
\begin{align}
    \rho & \sim |r - r_c|^{\beta_D}, \quad \text{with} \quad \beta_D = 1. \label{eq:betaD} \\
    \rho & \sim |r - r_T|^{\beta_T} \quad \text{with} \quad \beta_T = \frac{1}{2}. \label{eq:betaT}
\end{align}

The scaling laws in Eqs.~(\ref{eq:betaD}, \ref{eq:betaT}) imply distinct scaling regimes, consistent with Fig.~\ref{fig:log_allee_scaling}.

\subsection{External flux}
We add a small external flux term $w$ (e.g., migration) to the GAL map:
\begin{align}
    x_{t+1} = f(x_t) + w = A x_t^3 + B x_t^2 + C x_t + w \label{eq:gal_map_w}
\end{align}
The stationary condition is given by
\begin{align}
F(\rho, w) = A \rho^3 + B \rho^2 + (C - 1) \rho + w = 0.
\label{eq:steady_state_w}
\end{align}
At criticality ($C \to 1, B\neq0$) the dominant terms are  $B \rho^2 + w \approx 0$. Hence, the scaling   is
\begin{align}
\rho(r=r_c) \sim w^{1/\delta^w_D}, \quad \text{with} \quad \delta^w_D = 2. \label{eq:deltaD}
\end{align}
At tricriticality ($C \to 1, B \to 0$) the dominant terms are given by $A \rho^3 + w \approx 0$. Thus, we find
\begin{align}
\rho(r=r_T) \sim w^{1/\delta^w_T}, \quad \text{with} \quad \delta^w_T = 3. \label{eq:delta_wT}
\end{align}
These scaling relations reveal that the density $\rho$ responds differently to external perturbations at both the critical and tricritical points, as corroborated by  Fig.~\ref{fig:log_allee_scaling}.

\subsection{Dynamical susceptibility}
To further analyze the response of the stationary state to a small external flux $w$, we compute the dynamical susceptibility
\begin{align}
\chi \equiv   \pdv{\rho}{w} \ . \label{eq:chi_def}
\end{align}
Differentiating Eq.~\eqref{eq:steady_state_w} with respect to $w$, we obtain:
\begin{align}
\pdv{F}{w} = 
(3A\rho^2 + 2B\rho + (C - 1))\chi + 1 = 0. \label{eq:susceptibility_general}
\end{align}
When $C > 1$, the linear term  dominates in both the critical and tricritical cases. From Eq.~\eqref{eq:susceptibility_general}, we   can calculate the following scaling relations for the susceptibility:
\begin{align}
\chi &\sim |r - r_c|^{-\gamma_D} \quad \text{with} \quad \gamma_D = 1, \label{eq:gammaD}
\\
\chi &\sim |r - r_T|^{-\gamma_T} \quad \text{with} \quad \gamma_T = 1. \label{eq:gammaT}
\end{align}
These exponents reveal a similar scaling for the dynamical susceptibility, as  confirmed in Fig.~\ref{fig:log_allee_scaling}.

\subsection{Universal crossover function}
We now return to the case without an external flux. We define deviations from the TCP using Eq.~(\ref{eq:tcp_B_C_condition}):
\begin{align} 
\Delta C &= C - C_T = C - 1, \label{eq:DeltaC} \\ 
\Delta B &= B - B_T = B. \label{eq:DeltaB}
\end{align}

Substituting into Eq.~\eqref{eq:steady_state} we obtain $A \rho^2 + \Delta B \rho + \Delta C = 0$, then after rescaling we get
\begin{align}
Y^2 + Y - V = 0, \label{eq:Y_equation}
\end{align}
where the scaled order parameter $Y$ and scaled control parameter $V$ are given  by 
\begin{align} 
Y &= \rho\frac{A}{\Delta B},  \label{eq:Y}
\\
V &= -A \frac{\Delta C}{(\Delta B)^2}. \label{eq:V}
\end{align}
The  relevant solution $Y=\mathcal{F}(V)$  of Eq.~\eqref{eq:Y_equation} is given by
\begin{align}
\mathcal{F}(V) = 
\frac{1}{2}\left( -1 + (1 + 4V)^{1/2} \right)
. \label{eq:universal_function}
\end{align}
 
This function is universal since it does not depend directly on the parameters of the model, but only on the scaled variable $V$, as confirmed by the simulations shown in Fig.~\ref{fig:log_allee_scaling}(f).

\subsection{Crossover exponent}
Following~\cite{lubeck2006tricritical}, we now adopt a standard crossover scaling ansatz for the TDP:
\begin{align}
\rho \sim \Delta_{h}^{ \beta^{T}/\phi^{T} }\,\mathcal{H}\!\left(\Delta_{r}\,\Delta_{h}^{ -1/\phi^{T} }\right), \label{eq:scaling_rho_ansatz}
\end{align}
where $\Delta_{h}$ is the distance from TCP, $\Delta_{r}$ is the distance from the critical point, $\mathcal{H}$ is a universal crossover function, and $\phi^{T}$ is the crossover exponent.

Note that from Eq.~\eqref{eq:Y}-\eqref{eq:V} we have 
\begin{align} 
\rho &\sim   \Delta B \ \mathcal{F}( V),  \label{eq:Y2}
\\ V &\sim \Delta C \ (\Delta B)^{-2}. \label{eq:V2}
\end{align}

To apply the ansatz Eq.~\eqref{eq:scaling_rho_ansatz} to our universal function we must map $(\Delta B,\Delta C)$ to  $(\Delta_{h},\Delta_{r})$.
A natural mapping consistent with Eqs.~(\ref{eq:B_coeff}-\ref{eq:C_coeff}) is
$\Delta B \propto \Delta_{h}$ and 
$\Delta C \propto \Delta_{r}$.
 Thus, 
\begin{align} 
\rho &\sim   \Delta_h \ \mathcal{F}( \Delta_r \ (\Delta_h)^{-2})
\label{eq:scaling_rho_2}
\end{align}
Comparing Eqs.~(\ref{eq:scaling_rho_ansatz}, \ref{eq:scaling_rho_2}) we obtain $\beta^{T}/\phi^{T}  = 1$ and $-1/\phi^{T} = -2$. Therefore,
\begin{align}
\phi^{T} &= 1/2
\end{align}
This is the crossover exponent of the TDP universality class~\cite{lubeck2006tricritical,grassberger2006tricritical}.

\begin{figure}[!htb]
\centering
\includegraphics[width=0.456\textwidth]{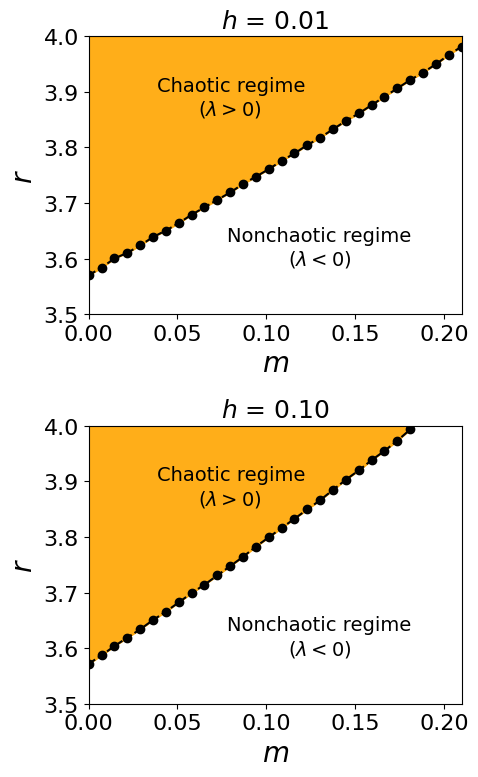}
\caption{Regime diagram in the GAL map as a function of the Allee parameter $m$. Inside the chaotic region there are stable windows, but this plot focuses on threshold $r_{\text{chaos}}$ where   
$\lambda$ first becomes positive. }
\label{fig:log_allee_onset_of_chaos}
\end{figure}

\subsection{Finite-time Lyapunov exponent}

To quantify the average separation of nearby trajectories 
we compute the finite-time Lyapunov exponent (FTLE) 
\begin{align}
\lambda_t \equiv \frac{1}{t}\sum_{i=0}^{t-1} \log\big|f'(x_i)\big| ,
\label{eq:lambda_t_def}
\end{align}
where for the GAL map we obtain 
\begin{align}
\lambda_t &= \frac{1}{t} \sum_{i=0}^{t-1} \ln |3A x_i^2 + 2B x_i + C|. \label{eq:lyapunov_gal}    
\end{align} 
For the critical case ($C \to 1, B\neq0$) we know that $x_t\sim t^{-1}$. Using the leading term of Eq.~\eqref{eq:lyapunov_gal} we obtain
\begin{align}
\lambda_t (r_c)  &\sim  \frac{1}{t}\sum_{i=1}^{t} 2B x_i
   \;\sim\; t^{-\theta_D}\,\log t, \quad \theta_D=1.
\label{eq:lambda_critical_scaling} 
\end{align}
For the tricritical case ($C \to 1, B \to 0$), we know that  $x_t\sim t^{-1/2}$, therefore
\begin{align}
\lambda_t (r_T)  &\sim  \frac{1}{t}\sum_{i=1}^{t} 3A x_i^2
\;\sim \;   t^{-\theta_T}\,\log t, \quad  \theta_T=1.
\label{eq:lambda_tricritical_scaling}
\end{align}
Both the critical and tricritical cases exhibit the same form $\lambda_t \sim t^{-1} \log t $,  as confirmed in 
Fig.~\ref{fig:log_allee_lyapunov}.

\subsection{Chaos}

In the long-time limit, we compute the Lyapunov exponent  
$\lambda = \lim_{T \to \infty} \lambda_T$   
where $\lambda_T$ is the FTLE, Eq.\ref{eq:lambda_t_def}.

In Fig.~\ref{fig:log_allee_bifur_lyap}, it is clear how the GAL dynamics changes with increasing $r$. 
The transition from a fixed-point to chaos is mediated by the Feigenbaum-like scenario (cascade of period-doubling bifurcations) even in the presence of the Allee-effect ($m>0$). 

In Fig.~\ref{fig:log_allee_onset_of_chaos}, we see how the
magnitude of the Allee effect, $m$, modulates the transition to chaos in the GAL map. While the standard logistic map ($m=0$) becomes chaotic at
$r\approx 3.5699$ (accumulation point), increasing $m$ systematically raises this threshold. That is, chaotic regimes are less likely to occur when  the Allee effect is present, which is a result consistent with the previous literature~\cite{fowler2002population,nath2022refugia}.

\section{Final remarks}\label{sec:finalremarks}

We introduced the generalized Allee-logistic (GAL) map, which is a 
model that provides a new building block for the interface between critical and tricritical phenomena, Allee effects, and chaotic maps, extending previous related contributions
~\cite{feigenbaum1978quantitative,chang1981iterative,kapustina2001scaling,kuznetsov2001two,ambika2002critical,xu2003experimental,kuznetsov2006effect,stynes2010scaling,teixeira2015convergence,girardi2019comment,corral2018finite,corral2025universal,martin2025finite,polli2025power,aleixo2009populational,leonel2019allee,el2023allee}.  
Our exact analytical calculations and simulations reveal a rich phase diagram governed by an exact TCP and universal crossover function. 
 Note that our GAL map is different from other generalizations of the logistic family~\cite{radwan2013some,borujeni2015modified,lawnik2017generalized,da2017route,sayed2017generalized,hamada2025investigating,zhang2025chaos,abdellah2025generalized,pires2025composing,chatterjee2023attack,gao2025chaos,vijayan2025universality,edelman2025fractional,pawela2026matrix}.

The scaling results (summarized in Table~\ref{tab:comparison_DP_TDP}) remarkably agree with MF-DP  and MF-TDP universality classes~\cite{lubeck2006tricritical,grassberger2006tricritical}, thereby expanding the set of models in the TDP class~\cite{cellai2011tricritical,dias2014adsorbed,araujo2015kinetic,hnativc2019tricritical,jo2020tricritical,pizzi2021bistability,pires2023tricritical,gao2026tricritical,gao2026tricritical}. This correspondence is consistent with the Janssen-Grassberger conjecture~\cite{janssen1981nonequilibrium,grassberger1982} for absorbing-state transitions~\cite{hinrichsen2000non,odor2004universality}. Clearly, this universality equivalence refers to scaling exponents and not to the microscopic dynamics since our GAL map is not stochastic.  This is  aligned with the   essence of universality: similar global features for different microscopic rules.

From an ecological perspective, our results highlight that the Allee effect plays a dual role. On the one hand, it induces a discontinuous transition to extinction (which is absent in the standard logistic map); on the other hand, increasing the Allee strength can delay the onset of chaos. 

Future work could explore the effects of additive and  multiplicative noise on the GAL map, which may reveal an even richer phenomenology.
\section*{Acknowledgments} 
We gratefully acknowledge CNPq, CAPES, FUNCAP and the National Institute of Science and Technology for Complex Systems in Brazil for financial support.

\bibliography{main.bib}

\end{document}